\documentclass[aps,preprint,amssymb,12pt,floatfix]{revtex4}
\usepackage{subfigure}
\usepackage{graphicx}
\usepackage{lscape,graphicx}
\usepackage{rotating}
\usepackage{epstopdf}
\usepackage{color}    
\usepackage{amsmath,amsthm}

\newcommand{\dprod}{\displaystyle\prod}

\begin{document}

\title{Random First Order Phase Transition Theory of the Structural Glass Transition}
\author{T. R. Kirkpatrick$^{1,2}$ and D. Thirumalai$^{1,3}$}
\affiliation{$^1$Institute for Physical Science and Technology, University of Maryland, College Park, MD 20742\\
$^2$Department of Physics, University of Maryland, College Park, MD 20742\\
$^3$Department of Chemistry and Biochemsitry, University of Maryland, College Park, Maryland 20742, USA}
\date{\today}

\begin{abstract}

We describe our perspective on the Structural Glass Transition (SGT) problem built on the premise that a viable theory must provide a consistent picture of the dynamics and statics, which are manifested by large increase in shear viscosity and thermodynamic anamolies respectively. For the static and dynamic description to be consistent we discovered, using a density functional description without explicit inclusion of quenched random interactions and a mean-field theory,  that there be an exponentially large number of metastable states at temperatures less than a critical transition temperature, $T_A$. At a lower temperature ($T_K < T_A$), which can be associated with the Kauzmann temperature, the number of glassy states is non-extensive.  Based on this theory we formulated an entropic droplet picture to describe transport in finite dimensions in the temperature range $T_K < T < T_A$.  From the finding that glasses are trapped in one of many metastable states below $T_A$ we argue that during the SGT law of large numbers is violated. As a consequence in glasses there are sub sample to sub sample fluctuations provided the system is observed for times longer than the typical relaxation time in a liquid.  These considerations, which find support in computer simulations and experiments, also link the notion of dynamic heterogeneity to the violation of law of large numbers. Thus, the finding that there is an extensive number of metastable states in the range $T_K < T < T_A$ offers a coherent explanation of many of the universal features of glass forming materials. 
\end{abstract}

\maketitle

\section{I. INTRODUCTION}

The Structural Glass Transition (SGT) in a variety of materials whose molecular constituents are chemically different exhibits several universal characteristics.  This has prompted a quest for describing a universal mechanism of  the nature of the glass transition. Over twenty years ago there were a number of significant contributions to
the understanding of the SGT. They included the mode
coupling theory of the glass transition \cite{Leutheusser84PRA,Gotze92RepProgPhys,Kirkpatrick85PRA}, the random first order transition
theory of the glass transition \cite{Kirk89JPhysA,Kirk89PRA}, and its connection to dynamic theories, and
to a lesser extent, the kinetically constrained kinetic models of the glass
transition \cite{Fredrickson84PRL}. Much of this work was, in turn, motivated by research done ten
years earlier on the spin glass problem \cite{Edwards75JPhysF,Sherrington75PRL,MezardBook}.

In this short perspective we discuss a theory for the SGT
 that is based on using frozen density fluctuations as an order
parameter to characterize the SGT \cite{Kirk89JPhysA}. The particular type of transition that
arises from our considerations has been referred to as a random first order phase
transition (RFOPT) \cite{Kirk89JPhysA,Kirk89PRA}. Originally this sort of phase transition was
theoretically found in certain class of exactly soluble spin glass models \cite{Kirk87PRL,Kirk87PRB,Kirk87PRBa}, but it was
subsequently realized that it can naturally occur in systems without
quenched disorder and in continuum field theory models \cite{Kirk89JPhysA}. In part we will
focus on the connection between a static approach to describe the SGT and a
dynamical theory, notion that was first emphasized in \cite{Kirk87PRA}. Too often these approaches are presented as being
distinct from each other while a careful study shows that they are, in fact,
closely related.  Indeed, establish such a connection is crucial to understanding the nature of the SGT\cite{Jackle86RepProgPhys,Martinez01Nature,Angell95Science}. In  linking the static and dynamic theories of the glass transition we discovered that the universal features of glasses are 
characterized by the nature of an extensive number of metastable states that emerge below a characteristic temperature \cite{Kirk89JPhysA}.  

Physically there are obvious similarities between spin and structural
glasses \cite{Kirk95TTSP}. A structural glass is a frozen liquid: A snap shot of a structural
glass looks identical to a snapshot of a liquid, i.e., in both cases there
is no long range spatial order. It is only when the system develops in time
that a glass is obviously different than a liquid since in the former case
the molecules are localized to a small region of space on experimental time
scales while in the latter case, any molecule diffuses arbitrary far from
where it started as $t\rightarrow \infty $. Similarly, a spin glass is a
frozen paramagnet: A snapshot of a SG looks identical to a snapshot of a
paramagnet, i.e., there is no long ranged magnetic order. When the systems
evolves in time the local magnetic moment or spin points in a specific
(time) averaged direction in the SG phase while in the paramagnetic phase
the spin direction randomly fluctuates in time so that the time averaged
magnetic moment is zero. In both the liquid and spin problems it is clear
that there is at least an effective phase transition into the glassy state,
and that there is broken ergodicity at the glass transition since time
averages no longer equal full ensemble averages. Moreover, there are violations of the fluctuation-dissipation relations thus making the dynamics dependent on initial conditions
\cite{Thirum88PRB,Cugliandolo93PRL}, and aging effects become relevant \cite{Cugliandolo07PhysicaA}.

For some time it was thought that there were important conceptional
differences between structural and spin glasses. In particular, in SG
problems an important input is that there is quenched disorder. In
structural glasses, on the other hand, there is quenched disorder only in
the glassy phase, and it is self-generated. In other words, the Hamiltonians of glass forming materials are not
random whereas the emergence of glassy behavior in SG is due to the presence of quenched random interactions between the spins. We now know that this difference
is not so relevant. In fact, the important contribution of \cite{Kirk89JPhysA} was that
it was the first paper to show that the methods that were developed to describe
RFOPT in SG models, which   appeared to be resticted to models with quenched
disorder could, in fact, be used in any system where there are many statistically distributed
metastable states \cite{Kirk89JPhysA}. This crucial discovery that allowed us to produce a consistent static and dynamical theory of the SGT using a density functional Hamiltonian has been subsequently elaborated and expanded by numerous authors (see \cite{Mezard09CondMat} and references therein).
 
The plan of this perspective is as follows. In Section II we introduce specific
static and dynamic density functional models for the SGT. In Section III the
static approach to understanding the SGT is described. In particular, we use
a combination of mean-field like approximations and a replica approach to
deal with the self-generated randomness in structural glasses to conclude
there is a special temperature, conventionally denoted by $T_{A},$ below
which there are an exponentially large number of glassy solutions. In
Section IV we show the purely dynamical approach yields results for the SGT
that are identical to the static approach. In Section IV we describe a
related scaling and droplet approach to RFOPT and to the STG transition in
particular. We also discuss how to characterize the liquid system below $%
T_{A}$. We see that a Kauzmann temperature, $T_{K}$, is naturally present in
this approach, and that a true glass transition is possible if $T_{K}$ could
be reached while maintaining the system in equilibrium. In this Section we
also review some speculation on transport at $T_{K}$ is approached.  A consequence of the droplet picture is that
the law of large numbers is violated in the glassy phase. The implication is that when observed over a period of time various properties in glasses vary from region to region, which
naturally explains the emergence of dynamical heterogeneity, and broken ergodicity.  We
conclude in Section VI with a discussion.

\bigskip

\section{DENSITY FUNCTIONAL MODELS FOR THE GLASS TRANSITION}

A glass can be characterzied as an amorphous that is described in terms of statistically distributed
density field, or as a dynamically frozen density fluctuation. This
motivates using the number density, $n(\mathbf{x},t),$as the order parameter
for the SGT where $(\mathbf{x},t)$ are space-time points. To be specific we
will give a very explicit functional field theory for $n$, and dynamical
equations for the density. Once this is done, we will discuss how to
characterize the glass transition by the behavior of $n$ as a punitive glass
transition is approached within these simple models. More general theories
can, of course, be considered.

\subsection{Density Functional Hamiltonian (DFH)}

The static model for the glass transition reviewed here was motivated by
results for spin glass models without reflection symmetry \cite{Kirk87PRL}. The resulting
phase transition is known as a random first order phase transition (RFOPT).
Originally it was thought that this sort of unusual transition required
the existence of quenched disorder as is the case in spin glass systems \cite{MezardBook}, but not
in systems undergoing a structural glass transition, where it is said that
the disorder is self-generated. As mentioned in the Introduction this turned
out not to the case, as we first showed in 1989 \cite{Kirk89JPhysA} (referred to from now on as KT). The theoretical ideas in KT form the basis of many subsequent developments in the SGT. Indeed, the generality of the conclusions in KT has subsequently been
established by others as well \cite{Franz95PRL}.

The model DFH is,%
\begin{eqnarray}
\beta \mathcal{H=}&-\mu \int d\bf{x}\delta n(\bf{x}+ \frac{1}{2}\int d \nonumber
\bf{x}_{1}d\bf{x}_{2}\delta n(\bf{x}_{1})\chi^{-1}(\bf{x}%
_{1}\bf{-x}_{2})\delta n(\bf{x}_{2})+ \frac{g_{3}}{3}\int d\bf{%
x[\delta n(\bf{x})]^{3}}\\
&+\frac{g_{4}}{4}\int d\bf{x}\bf{[}\delta n(\bf{x})\bf{]}^{4} -\int d\bf{x}H(\bf{x})\delta n(\bf{x}) 
\end{eqnarray}%
where $\mu $ is the chemical potential, $H(\mathbf{x})$, is a small ($%
\rightarrow 0$), random, symmetry breaking external field whose role will
become clear, and $g_{3}$ and $g_{4}$ are nonlinear coupling constants whose
magnitudes are chosen such that a systematic self consistent expansion in
density fluctuations is possible. The wavenumber($k$)-dependent $\chi (k)$ is related to
the static structure factor and contains information on the short range
order in the fluid.

\subsection{Dynamical model}

Since we are characterizing the SGT in terms of frozen density fluctuations
we need a dynamical equation for space and time dependent density
fluctuation, $\delta n(\mathbf{x,}t).$ We chose conservative relaxational
dynamics, which reflects the fact that at a molecular length scale, density
fluctuations are diffusive. The dynamical equation is,%
\begin{equation}
\frac{1}{\Gamma _{0}}\partial _{t}\delta n(\mathbf{x,}t)=\nabla ^{2}\frac{%
\delta (\beta \mathcal{H})}{\delta (\delta n(\mathbf{x,}t))}+\xi (\mathbf{x,}%
t)  
\end{equation}%
with $\xi (\mathbf{x,}t)$ the usual Gaussian noise term and $\Gamma _{0}$ a
bare kinetic coefficient that sets the microscopic time scale. More general
dynamics including 'mode coupling' terms can also be considered.

\bigskip

\section{STATIC THEORY OF THE GLASS TRANSITION}

The static theory of the glass transition starts with the DFH given by Eq.
(1). We introduce two related key notions \cite{Kirk89JPhysA}. First, we imagine an order
parameter description in terms of frozen density fluctuations. Since the
glassy state is amorphous or aperiodic \cite{Singh95PRL} it is most naturally specified by a
probability measure $\mathcal{P[}\delta n\mathcal{]}$. Secondly, we allow
for a large number of macro states or pure states (at least on a given time
scale). These states are characterized as follows. Denote a particular
macroscopic state by the label s, with the density field in that state given
by $n_{s}=n_{o}+\delta n_{s}.$ We denote the free energy of this state by $%
F_{s}$. Next compute $F_{s}$ from Eq. (1) by standard loop expansion
techniques. We then allow for a possibly large number of statistically
similar but different states by using a partition function defined by,%
\begin{equation}
Z=\sum_{s}\exp (-\beta F_{s})=\int D[\delta n]\Delta (\delta n)\exp (-\beta
F)\dprod \left( \frac{\delta \beta F}{\delta (\delta n(%
\mathbf{x}))}\right) .  
\end{equation}%
Here $\Delta (\delta n)=\left\vert \det \delta ^{2}F/\delta n^{2}\right\vert 
$ normalizes the delta function in the equation given above. Equation (3) defines a
probability measure $\mathcal{P}[\delta n]$ for the field $\delta n$. In
usual phase transition problems $\mathcal{P[}\delta n\mathcal{]}$ is a delta
function at the unique (or, more generally, at all globally
symmetry-related) equilibrium state(s) of the system. However, in general,
Eq. (4) is capable of describing a large number of symmetry-unrelated
states that are statistically distributed.

We have used these equations to solve for the SGT as follows. We define the
correlation function,%
\begin{equation}
Q(\mathbf{x}_{1},\mathbf{x}_{2})=<\delta n(\mathbf{x}_{1})\delta n(\mathbf{x}%
_{2})>,  
\end{equation}%
and an analogous density response function, $R(\mathbf{x}_{1}\mathbf{,x}%
_{2}).$ Here the angular brackets denote an average with weight $\mathcal{P[}%
\delta n\mathcal{]}$. We solve for these two correlation functions using a
standard loop expansion. We assume that the field $H$ is a small Gaussian
random field which statistically breaks the symmetry of the liquid phase to
a glassy phase. The random field serves as an external coupling term
conjugate to the Edwards-Anderson order parameter, $Q$, which characterizes
the glassy phase. The variance of $H$ is set equal to zero at the end of the
calculation. Carrying out a self-consistent expansion, using the fact that $%
g_{3}$ and $g_{4}$ are small, yields closed equations for $R$ and $Q.$ In
wavenumber space these equations are,%
\begin{equation}
R(k)=n_{0}S(k)-Q(k)  
\end{equation}%
and%
\begin{equation}
Q(k)=R(k)\left( 2g_{3}^{2}\int_{k_{1}}Q(k-k_{1})Q(k_{1})\right) . 
\end{equation}%
An identical  nonlinear equation for the glassy order parameter is obtained in the next
Section using a dynamical approach. The solution of the resulting equation has been discussed elsewhere (\cite{Kirk89JPhysA} and references therein).  Here
we note that nontrivial $Q$ solutions become possible below a temperature, $%
T_{A}$, and at this tempearture $Q$ jumps discontinuously to a nonzero
value \cite{Kirk89JPhysA}. In giving Eqs.(5) and (6), we explicitly considered random solutions by
requiring $V^{-1}\int d\mathbf{x}\delta n(\mathbf{x})\rightarrow 0$, where $%
V $ is the volume, even though the square (spatial) average of $\delta n$ is
nonzero. These two conditions hence lead to the the moniker, random first
order phase transition.

We note that the manipulations leading to Eqs.(5) and (6) are similar to those used
for mean field spin glasses. Physically, the SGT theory is similar to mean
field SG theories because; (1) The term in the brackets in Eq.(6)
represents self-generated randomness, and (2) Mean field like approximations
were used in deriving Eq.(6). Also, in deriving Eqs.(5) and (6) we used an
infinitesimal Gaussian random field to set up the perturbation theory,
introduce replicas, and use a replica symmetry breaking scheme that assumed
only self-overlap of the metastable glassy states, as one does for SG
systems with RFOPT. We stress that at the end of our calculation we set $H=0$%
, so that there is no 'quenched' randomness. We note parenthetically that this method of locating a particular pure state has been used in the STG
problem studied by replica methods \cite{Monasson95PRL}

We next discuss the physical significance of the transition temperature
where the Eqs.(6) first has a nontrivial solution, $T_{A}$. We define
two, in general, distinct free energies,%
\begin{equation}
F_{c}=\sum_{s}\exp (-\beta F_{s})  
\end{equation}%
and%
\begin{equation}
\overline{F}=\frac{\sum_{s}F_{s}\exp (-\beta F_{s})}{\sum_{s}\exp (-\beta
F_{s})}. 
\end{equation}%
where $F_{c}$ is the usual canonical free energy while $\overline{F}$ is the
component weighted free energy. Direct calculation in the glassy state gives 
$F_{c}$ =$F_{L}$, with $F_{L}$ the liquid state free energy that does not
depend on $Q(k)$, and $\overline{F}>F_{c}$. The inequality $F_{c}\neq 
\overline{F}$ can occur if and only if the states leading to Eqs.(5) and (6) are
metastable and there are an infinite number of such states. Since $%
\overline{F}>F_{c}$ (see KT for additional discussions), it follows that at $T_{A}$ the liquid, within our
mean-field like approximations, freezes into a metastable glass that is
stabilized by an exponentially large solution degeneracy, i.e., there is a
complexity, or state entropy associated with the states below $%
T_{A}$. Mathematically this entropy, $S_{s}$, is defined by, 
\begin{equation}
TS_{s}=\overline{F}-F_{c}.  
\end{equation}%
Note that within our approximate calculations $F_{c}$ is not a physically
meaningful free energy because it contains an entropic term that is a
measure of states not probed in the time scales in which our calculations
are valid.

In Section V we argue that $T_{A}$, is not the glass transition temperature.
Rather, beacuse of the appearance of many glassy states, it is the
temperature below which activated dynamics play an important role and the
dynamics become increasingly sluggish. There is a lower Kauzmann
temperature, $T_{K}$, where the number of glassy states become nonextensive
and where there is a true glass transition.

\bigskip

\section{DYNAMICAL THEORY OF THE GLASS TRANSITION}

The dynamical theory of the SGT starts with Eqs. (1) and (2). We
consider the density time correlation function,%
\begin{equation}
C(\mathbf{k,}t)(2\pi )^{d}\delta (\mathbf{k+k}^{\prime })=<\delta n(\mathbf{%
k,}t)\delta n(\mathbf{k}^{\prime },0)>.  
\end{equation}%
The glassy state is defined by frozen density fluctuations or by a non-zero
Edwards-Anderson order parameter \cite{Edwards75JPhysF},%
\begin{equation}
q(k)=q_{EA}(k)=\lim_{t\rightarrow \infty }C(\mathbf{k},t),  
\end{equation}%
and we assume that the glassy state is statistically homogeneous and
isotropic. The fact that the glassy state has the same statistical
properties as the liquid state is necessary to establish the connection
between the static and dynamic approaches to the SGT.

Treating the nonlinear terms in Eq. (1) as small, the self-consistent
one-loop approximation for $\widehat{C}(\mathbf{k,}\omega )$, the one sided
Fourier transform of $C(\mathbf{k,}t),$is, 
\begin{equation}
\widehat{C}(\mathbf{k},\omega )=C(\mathbf{k},t=0)[-i\omega +\Gamma
_{R}(k,\omega )]^{-1},  
\end{equation}%
with $C(\mathbf{k,}t=0)=n_{0}S(k)$ the static structure factor and $\Gamma
_{R}(k,\omega )$ a renormalized kinetic coefficient,%
\begin{equation}
\Gamma _{R}^{-1}(k,\omega )=\frac{1}{\Gamma _{0}k^{2}}+2g_{3}^{2}\int_{%
\mathbf{k}_{1}}\int_{0}^{\infty }dt\exp (-i\omega t)C(\mathbf{k-k}_{1}%
\mathbf{,}t)C(\mathbf{k}_{1},t).  
\end{equation}

Equations (12) and (13) are of a standard form of the equations derived in the
so-called mode coupling theory of the glass transition, although the
conceptual origin of Eq.(13) is different than in the original mode
coupling theory. Equation (13) predicts a continuous slowing down of
density fluctuations and a freezing at a temperature denoted by $T_{A}$ \cite{Gotze92RepProgPhys}. The
equation of state for the frozen density fluctuations is obtained by
inserting Eq. (11) in Eqs. (12 and 13) and obtaining,%
\begin{equation}
q(k)=n_{0}S(k)\frac{2g_{3}^{2}\int_{\mathbf{k}_{1}}q(\mathbf{k-k}_{1})q(%
\mathbf{k}_{1})}{1+2g_{3}^{2}\int_{\mathbf{k}_{1}}q(\mathbf{k-k}_{1})q(%
\mathbf{k}_{1})}.  
\end{equation}%
It is easy to show that Eq.(14) is identical to Eq. (6) if we identify $%
q(k)$ with $Q(k)$. 

To summarize, this dynamical approach leads to a continuous freezing of
density fluctuation and the frozen density fluctuations can be described by
either a static theory, or by the dynamical approach. The static approach
has the advantage that one can understand the freezing in terms of the
number of states etc. The freezing temperature is denoted by $T_{A}$ because
the self-consistent dynamical approach clearly ignore activated dynamics,
which dominate transport at low temperatures. This is, in turn, consistent
with the static approach where the freezing occurs into metastable glassy
states, which can only be precisely defined in some sort of mean-field limit
where activated dynamical processes do not occur. In the next Section the
temperature region $T<T_{A}$ is considered using non-perturbative scaling
and droplet ideas.

\bigskip

\section{SCALING AND DROPLET CONSIDERATIONS}

\subsection{\textbf{{\protect\normalsize {Activated transitions, entropic
droplets, and growing correlation length}}}}

The mean field theory based on precise calculations using a density
functional Hamiltonian without quenched disorder shows that in the
temperature range , $T_{A}<T<T_{K}$, the system is frozen in one of the
exponentially large number of metastable states. Flow from one of these
state (say $\alpha $) to another ($\gamma $) cannot be described within the
MFT because $\alpha $ and $\beta $ are two disjoint ergodic states. In order
to account for the observed non-Arrhenius slowing down of transport in
glassy systems, which is often captured in terms of the Vogel-Fulcher
equation, 
\begin{equation}
\tau (T)=\tau _{0}exp[{\frac{DT}{(T-T_{K})}}]  
\end{equation}%
Kirkpatrick, Thirumalai, and Wolynes  \cite{Kirk89PRA} (KTW) introduced a new scaling theory based on entropic driving forces, which were argued to be relevant for
transport \cite{Kirk87PRBa}. In Eq. (15) $\tau _{0}$ is a microscopic relaxation time at $%
T_{A}$ and $D$ is a positive constant. It follows from the KT theory described in Sections III and IV it
that the emergence of multiple metastable minima below $T_{A}$ can
be quantified in terms of the state entropy $S_{s}$, which is the difference
between the canonical and component averaged free energies. In the droplet
picture of activated transitions it is $S_{s}$, which is distinct from the
configurational entropy in the Adam-Gibbs theory \cite{Adam65JCP} (for a detailed discussion
see \cite{Kirk89PRA}), is the driving force for activated transition.
Consider a region in a glassy state of size $L^{d}$ and let us estimate the
probability of nucleating another glassy state inside $L^{d}$. The driving
force for being able nucleate a glassy state one inside the other has to be
entropic because the various glassy states have roughly the same free
energy. Because there are a vast number of accessible (on a long time scale)
glassy states such an entropically driven nucleation is possible. Within the
droplet picture the driving force for nucleation is $\sim Ts_{s}L^{d}$ where 
$s_{s}$ is the state entropy per unit volume. The formation of domain within a
domain is opposed by surface free energy cost, which can scale at most as $%
\sigma L^{(d-1)}$ where $\sigma $ is the surface tension (see below for a
careful treatment of the scale-dependence of the surface tension between two
distinct glassy domains). Balancing these two free energies gives the
typical size of the glassy cluster $L\ast \sim \frac{\sigma T}{s_{s}}$ and
the barrier to activated transport is $\Delta F\ast \sim (\frac{\sigma T}{%
s_{s}})^{(d-1)}$. We see that the entropic droplet theory naturally follows from considering the ramifications of the mean-field theory for finite dimensional systems.

The natural generalization of the MFT to describe activated transitions is
to assume that flow below $T_{A}$ is triggered by creation and destruction
of mosaic states within a large glassy cluster whose size $\xi \sim t^{-\nu }$ ($%
t=\frac{(T-T_{K})}{T_{K}}$) diverges at $T_{K}$. The entropic droplet
picture, that was inspired by the MFT and fluctuation theory \cite{Kirk89PRA}, has been used
to show $\nu =\frac{2}{d}$. The time scale associated with these processes
increases as the temperature decreases below $T_{A}$ eventually diverging at 
$T_{K}$ as described in Eq. (15). As long as the size of the glassy domain is
large then the entropic driving forces are opposed by surface free energy
cost that scales as 
\begin{equation}
F_{opposing}\approx \gamma L^{\theta }  
\end{equation}%
where $\theta \leq \frac{d}{2}$ if $\nu =\frac{2}{d}$. In terms of $s_{s}$
or equivalently $t$ (assuming $s_{s}$ varies linearly with $T$ close to $%
T_{K}$) the entropic driving force for activated transitions is 
\begin{equation}
F_{driving}\sim -At^{-(\nu d-1)}.  
\end{equation}%
Similiarly, $F_{opposing}\sim \gamma t^{-\nu \theta }$. The instability of
the droplets at large length scale requires that the exponent characterizing
the growth of the surface free energy be bounded by $\theta \leq \frac{(\nu
d-1)}{\nu }$.

These considerations can be used to describe the temperature-dependent
relaxation time near $T_{K}$. If the typical size of the glassy cluster
grows as $L\sim \xi \sim t^{-\nu }\sim t^{-\frac{2}{d}}$ then the typical
free energy barrier behaves as 
\begin{equation}
\Delta F\ast \sim t^{-(\nu d-1)}\sim t^{-1}  
\end{equation}%
which immediately results in the Vogel-Fulcher law (Eq. (15)). Notice in
order to obtain the Vogel-Fulcher equation the free energy cost opposing activated transport
must scale as 
\begin{equation}
F_{opposing}\sim \gamma L^{\frac{d}{2}}  
\end{equation}%
A more refined treatment that relies on a generalization of Villain's
conjectures  \cite{Villain85JPhys} for the Random Field Ising Model indeed shows that a
scale-dependent surface tension which vanishes on length scales greater than 
$\xi $ shows that Eq. (19) is indeed obeyed in the vicinity of $T_{K}$ \cite{Kirk89PRA}.
Although the KTW scaling picture offers a consistent picture of activated
transport that is wholly inspired by the precise theory described in \cite{Kirk89JPhysA}, it
still remains heuristic.

\subsection{\textbf{{\protect\normalsize {Violation of Law of Large
Numbers, Dynamical Heterogeneity, and loss of ergodicity}}}}

The precise connection between the static and dynamical description of the
glass transition made using the DFT (see Sections III and IV), which requires
the existence of an exponentially large number of metastable states below $%
T_A$, has been demonstrated by taking long time limits of the order
parameter $C(k,t)$. The long time (or more precisely on $t \ge \tau_c$ where 
$\tau_c$ is a correlation time) limit of $C(k,t)$ is zero in a liquid but
persists in glasses. In other words, only when the system develops in time
is there an obvious difference between liquids and glasses. In a liquid a
given molecule diffuses arbitrarily far from where it started as $t$ grows
whereas it is localized in space to a small region. The DFT calculations
show that the dynamical treatment and the static treatments give rise to
identical physical picture that is manifested by the appearance of a
non-zero glassy state Edwards-Anderson order parameter. Thus, within the
framework of RFOPT it is only by examining the link between spatial and time
correlations can the distinction between liquids and glasses be made.

As argued in the previous section it is fruitful to picture a glass as being
partitioned into mosaic states whose characteristic temperature-dependent
size is denoted by $\xi_{i}(T)$ where for the sake of generality we consider
variations in the sizes of the glassy clusters. Within the $\xi_{i}(T)$
structural rearrangements can be rapid but global relaxation requires
activated transitions, which involves nucleating new domains. Slow
structural fluctuations in glassy occurs because of entropic driving forces
that enable creation and formation of new glassy clusters by fluctuation
effects. As the degree of supercooling increases not only does $\xi_{i}(T)$
grow but also the relaxation time associated with particles that cross domain boundaries
exceeds the observation time scale ($\tau_{obs})$, thus leading to broken
ergodicity. These physical considerations that are embedded in the droplet
picture of the RFOPT also lead to violation of law of large numbers as the
STG occurs, which we illustrate by using the following arguments. Just as in
the scaling theory, we picture the glassy phase as being partitioned into
regions with size $\approx \xi_i(T)$ with $(\frac{\xi_i(T)}{a})^3$ being
sufficiently large that meaningful average over the number of particle
within $\xi_{i}(T)$ can be carried out. In the liquid phase ($T > T_A$) the
statistical properties of the liquid (for example the distribution of
energies of the particles in a glass forming system $P(\epsilon;
t|\xi_{i}(T))$ in the sub sample is {\it independent of $i$} and will coincide
with that of the entire sample provided $\xi_{i}(T)$ is large enough and $t
> \tau_c < \tau_{obs}$. This is a consequence of the law of large numbers.
In contrast, in the glassy phase each $\xi_i(T)$, which in the MFT
corresponds roughly to one of the frozen metastable states, is distinct, and
consequently each $P(\epsilon; t|\xi_i(T))$ can be distinct and will depend
on $i$. Thus, no single sub sample can characterize the distribution of
energies of the entire sample. In other words, in the glassy phase the law
of large numbers is violated, and there are sub sample to sub sample
fluctuations. Only by examining the \textit{entire sample} on $\tau_{obs}
> \tau(T)$ is ergodicity restored. We see that the so called dynamical
heterogeneity, which has been a characteristic of glass forming systems \cite{Sillescu99JNCS,Glotzer00JNCS,Donati99PRL} is
seen to be a consequence of the emergence of glassy clusters with the
characteristic sizes $\xi(T)$. Because of the variations in both
equilibrium and relaxation properties from sub sample to sub sample a glassy
phase is inherently heterogeneous.

The preceding arguments were illustrated using simulations of soft-sphere binary
mixtures in which the sample was divided into a
number of sub samples \cite{Thirumalai89PRA}. In the liquid phase $P(\epsilon ;t|\xi _{i}(T))$
coincides with the entire sample for all $i$ as long as $t>\tau _{c}$. In
contrast there are considerable variations in $P(\epsilon ;t|\xi _{i}(T))$
and are fragments of the entire sample. Thus, the dynamical heterogeneity is
really a consequence of the law of large numbers, and very much supports the
droplet scenario for activated transition set within the RFOPT context. A
corollary of the violation of the law of large numbers is that particles of
a specific type (say a large particle in a binary LJ mixture) belonging to
two distinct subsamples are not "statistically equivalent" even when $\tau
_{obs} \gg \tau _{c}$. This is in contrast to the liquid phase where on $%
t\approx \tau _{c}$ all particles of a given type are statistically
equivalent. Such a loss in statistical symmetry in the SGT is a time
averaged property and can only be inferred by examining the time evolution
of the system. The arguments and simulations reported by us \cite{Thirumalai89PRA} clearly showed
that dynamical heterogeneity and broken ergodicity naturally follow from  violation of the law of
large numbers.

Another consequence of the statistical inequivalence of any two subsamples (whose sizes are on the order of a typical $\xi(T)$) in the glassy phase is that ergodicity is broken in the SGT. To illustrate the concept of ergodicity breaking we introduce a measure referred to as the energy metric, $d(t)$, which is defined as
\begin{equation}
Nd(t) = \sum_{i=1}^{N} [\epsilon_i(t|R_{\alpha}(t)) - \epsilon_i(t|R_{\beta}(t)]^2
\end{equation}
where $\epsilon_i(t|R_{\alpha}(t)) = \frac{1}{t} \int_0^{t} ds E_i(s|R_{\alpha}(t))$. Here, $E_i(s|R_{\alpha}(t))$ is the energy of the $i^{th}$ particle at time $s$ and $R_{\alpha}(t)$ refers to a set of positions of the particles whose initial condition is labeled $\alpha$. Similarly, $\epsilon_i(t|R_{\beta}(t))$ is the corresponding quantity for the trajectory $\beta$.  If the system is ergodic on the time scale $\tau_{obs}$ then $d(t)$ vanishes as $t \rightarrow \tau_{obs}$, and therefore  $\epsilon_i(\tau_{obs}|R_{\beta}(\tau_{obs})) = \epsilon_i(t|R_{\beta}(t))$ {\it independent} of $alpha$ or $\beta$. This is the situation that pertains to the liquid phase. However, if ergodicity is broken, as is expected at the STG,  $d(t) \sim C$ ($C$ is a constant) suggesting that the two initial states do not mix on the time scale $\tau_{obs}$. As argued above it is the development in time rather than any equal time correlation functions that distinguishes a glass from a liquid.  It can be shown, using scaling-type arguments, that $\frac{d(0)}{d(t)} \approx D_E t$ where the "diffusion" constant $D_E$ is not unrelated to relaxation time set by the shear viscosity \cite{Thirumalai93PRE}. Thus, $N \frac{d(0)}{d(t)}$, which is extensive in $N$ and $\tau_{obs}$ in the liquid phase, remains only extensive in $N$ in the glassy phase because $\tau_(T) \gg \tau_{obs}$. We demonstrated these ideas using molecular dynamics simulations of two component softly repelling spheres as well Lennard-Jones  mixtures with additive diameters chosen to avoid crystallization.  At temperatures that are greater than $T_A$ we showed that  $\frac{d(0)}{d(t)}$ grows linearly as $t$ increases whereas it saturates in the glassy phase due to the inability to explore distinct regions of the configuration space.  The illustrations summarized here, which have been demonstrated by others using different language, follow directly from the physical picture that in the SGT the glass forming system is frozen into one of many disjoint ergodic states that do not mix (or become statistically equivalent) on $\tau_{obs}$.

\bigskip

\section{DISCUSSION}

The fundamental goal of any theory of glass forming materials should be to explain both the dramatic viscosity increase and thermodynamic anamolies starting from a theory appropriate for liquids. At the laboratory glass transition temperature $T_g$  the relaxation times far exceed the observation times and the heat capacity has a discontinuity suggesting that providing a kinetic description alone is insufficient. In addition, the goal of any theory of glasses must ultimately be described using quantities that can be measured in experiments. This perspective  presents a coherent theory that was advanced by us over twenty years ago, and which was guided by the goals outlined above. The theory and its implications for activated transitions, violation of law of large numbers and the related dynamical heterogeneity, and ergodicity breaking treats both the dynamical and static properties of glasses on equal footing. The major conceptual basis, which was discovered using a density functional description of glasses without quenched disorder, is that at $T \le T_A$ the system is frozen into one of many metastable states. In practical terms $T_A$ ($>T_g$) corresponds to a temperature at which $\eta \sim$ (1-10) poise. Such states are described by frozen density fluctuations from which emerges an Edwards-Anderson order parameter can be obtained from a purely static or a dynamical theory \cite{Kirk89JPhysA}.  

There are immediate consequences of the RFOPT of glass transition when applied to finite dimensions. Unlike in the mean field picture the metastable states are not disjoint and transport becomes possible on time scales comparable to $\tau(T)$, which of course, becomes exceedingly long as $T$ decreases. In the temperature range $T_K < T <T_A$ it is fruitful to think of glasses as being composed of  a large number of mosaic states on scales on the order of $\xi(T)$. From this picture we draw several significant conclusions.
\begin{enumerate}
\item

Transport in the temperature range, $T_K < T <T_A$,  is driven by activated processes the driving force for which are entropic in nature. Because the entropy vanishes linearly near $T_K$ it follows from our picture that the size of the domains must grow as $\xi \sim (T-T_K)^{-{\frac{2}{d}}}$.  The droplet theory \cite{Kirk89PRA},  constructed by  balancing  the entropic driving force and the opposing cost of creating an interface between two glassy states readily leads to the Vogel-Fulcher equation (Eq. 15. It is useful to comment on the typical values of $\xi(T_g)$ found in  practice,  Computer simulations of LJ mixtures \cite{Mountain92PRA} and colloidal glasses composed of mixtures of micron size charged particles \cite{Rosenberg89JPhysCondMatt} conclude that $t \sim 0.6$ which was used to show that $\xi(T_g) \approx  3 \sigma$ where $\sigma$ is the particle diameter. From the extracted values of $t \sim 0.1$ in several experiments \cite{Mohanty90JCP} we predict using the $t^{-\frac{2}{d}}$ scaling that $\xi(T_g) \approx 10 \sigma$ \cite{Berthier05Science}.  On these length scales there are in excess of fifty particles so that the activation barrier for transport is large enough that considerations from the scaling theory are appropriate.  

\item
The partitioning of a  glassy state into mosaic states with growing domain size suggests that law of large numbers must be violated, especially at temperatures less than $T_A$ \cite{Thirumalai89PRA}.  This implies that, when observed over a period of time that exceeds $\tau_c$ but is comparable to $\tau_{obs}$, any two mosaic states are statistically inequivalent.  As a consequence, glass is dynamically heterogeneous which implies that the statistical properties (averaged over a period of time greater than $\tau_C$) vary from one mosaic state to another. This is not the case in a liquid. These expectations are borne out in computer simulations. The conceptual basis of the origin of dynamic heterogeneity is intimately linked to the violation of law of large numbers \cite{Thirumalai89PRA}.

\item
Because of the statistical inequivalence of mosaic states on time scales comparable to $\tau_{obs}$ ergodicity is broken in the STG.  This is manifested in the ergodic measure, which is extensive in $\tau_{obs}$ in the liquid phase but becomes essentially independent of $\tau_{obs}$ in the glassy phase \cite{Thirumalai89PRA}. 

\end{enumerate}

\bigskip
 
{\bf ACKNOWLEDGMENTS}: We are grateful to grants from the National Science Foundation (DMR09-01907 and CHE09-14033) for support of this work.

\clearpage

\renewcommand{\baselinestretch}{1} 
\bibliography{Glasses}
\bibliographystyle{unsrt}

\end{document}